\documentclass[journal=nalefd,manuscript=letter]{achemso}
\usepackage{epsf}
\usepackage{amsfonts}
\usepackage[fleqn]{amsmath}
\usepackage{amssymb,yfonts,mathrsfs,bbm}
\usepackage{graphicx}
\usepackage{multirow}
\usepackage{longtable}
\usepackage{amsmath}

\title{Monolayer Molybdenum Disulfide Nanoribbons with High Optical Anisotropy}

\author{Jiang-Bin Wu}
\altaffiliation{Contributed equally to this work}
\affiliation{State Key Laboratory of Superlattices and Microstructures, Institute of Semiconductors, Chinese Academy of Sciences, Beijing 100083, China}

\author{Huan Zhao}
\altaffiliation{Contributed equally to this work}
\affiliation{Ming Hsieh Department of Electrical Engineering, University of Southern California, Los Angeles, CA 90089, USA}

\author{Yuanrui Li}
\altaffiliation{Contributed equally to this work}
\affiliation{Ming Hsieh Department of Electrical Engineering, University of Southern California, Los Angeles, CA 90089, USA}

\author{Douglas Ohlberg}
\affiliation{Intelligent infrastructure Lab, HP Labs, Hewlett-Packard Co., Palo Alto, California 94304, USA}

\author{Wei Shi}
\affiliation{State Key Laboratory of Superlattices and Microstructures, Institute of Semiconductors, Chinese Academy of Sciences, Beijing 100083, China}

\author{Wei Wu}
\email{wu.w@usc.edu}
\affiliation{Ming Hsieh Department of Electrical Engineering, University of Southern California, Los Angeles, CA 90089, USA}

\author{Han Wang}
\email{han.wang.4@usc.edu}
\affiliation{Ming Hsieh Department of Electrical Engineering, University of Southern California, Los Angeles, CA 90089, USA}

\author{Ping-Heng Tan}
\email{phtan@semi.ac.cn}
\affiliation{State Key Laboratory of Superlattices and Microstructures, Institute of Semiconductors, Chinese Academy of Sciences, Beijing 100083, China}

\begin{document}
\begin{abstract}
{Two-dimensional Molybdenum Disulfide (MoS$_2$) has shown promising prospects for the next generation electronics and optoelectronics devices. The monolayer MoS$_2$ can be patterned into quasi-one-dimensional anisotropic MoS$_2$ nanoribbons (MNRs), in which theoretical calculations have predicted novel properties. However, little work has been carried out in the experimental exploration of MNRs with a width of less than 20 nm where the geometrical confinement can lead to interesting phenomenon. Here, we prepared MNRs with width between 5 nm to 15 nm by direct helium ion beam milling. High optical anisotropy of these MNRs is revealed by the systematic study of optical contrast and Raman spectroscopy. The Raman modes in MNRs show strong polarization dependence. Besides that the $E'$ and $A'$$_1$ peaks are broadened by the phonon-confinement effect, the modes corresponding to singularities of vibrational density of states are activated by edges. The peculiar polarization behavior of Raman modes can be explained by the anisotropy of light absorption in MNRs, which is evidenced by the polarized optical contrast. The study opens the possibility to explore quasi-one-dimensional materials with high optical anisotropy from isotropic 2D family of transition metal dichalcogenides.}

\end{abstract}
{\bf Keywords}: Molybdenum Disulfide, Nanoribbon, Raman spectroscopy, Anisotropy.
\newpage


\maketitle

Two-dimensional (2D) transition metal dichalcogenide (TMD) layered materials such as MoS$_2$ have attracted great interest, due to its remarkable physical properties and excellent device prospect.\cite{geim2013van,lee2013synthesis,xu2014spin,xia_2014_two,zhao_2015_two,mak-2010-atomically,zeng-2012-valley,cao-2012-valley,binding-MoS2-prb,
mos2-trions-nm,fuhrer-2013-measurement,MIT-MoS2-nm} On the other hand, geometry of 2D materials plays a critical role in further tuning their properties. By patterning 2D materials into narrow (sub-20 nm) ribbons\cite{halfmetall-GNR-Nature,chemically-GNR-Science}, interesting phenomenon and novel properties can arise. A variety of properties of MoS$_2$ nanoribbons (MNRs) have been calculated by theory\cite{mos2-NR-jacs}. Monolayer (1L) zigzag MoS$_2$ nanoribbons (Z-MNRs) is predicted with the ferromagnetic and metallic behavior, irrespective of the ribbon width and thickness, and 1L armchair MoS$_2$ nanoribbons (A-MNRs) are nonmagnetic and semiconducting, and the band gaps converge to a constant value of 0.56 eV as the ribbon width increases.\cite{mos2-NR-jacs} Moreover, these properties of MNRs can be tuned by strain and electric field.\cite{Polarity-NR-jacs,tuning-zigzagmos2-jpcl,Electric-mnr-acsnano,strain-MNR-JPCC,defects-MNR-jpcc} However, the experimental study of the MNRs is still very limited due to the many challenges in patterning high quality sub-20 nm 2D material ribbon.\cite{nc-MNR-favrication} Here, we prepared MNRs with different widths between 5 nm to 15 nm using direct Helium ion beam milling. High optical anisotropy of MNRs is revealed by the systematic study of optical contrast and Raman spectroscopy on MNRs. The Raman modes in MNRs show strong polarization dependence. Besides the $E'$ and $A'$$_1$ peaks are broadened by the phonon-confinement effect, there are additional emerging modes activated by the edges. The polarization behavior of Raman modes is due to the anisotropy of the light absorption in the nanoribbons that is a result of the one-dimensional quantum confinement in patterned MNRs. The polarized optical contrast gives the evidence of such anisotropy in the nanoribbons.

Raman spectroscopy is one of the most widely used measurement technique that can reveal rich characteristics of 2D TMDs.\cite{Zhang_2015_csr} In bulk MoS$_2$, there are two prominent modes, ${E}_{2g}$ (383.6 cm$^{-1}$) and ${A}_{1g}$ (408.7 cm$^{-1}$), which correspond to two sulfur atoms and molybdenum atom of the same layer moving in the opposite in-plane direction and two sulfur atoms of the same layer moving in the opposite direction out-of-plane respectively. The ${E}_{2g}$ and ${A}_{1g}$ modes reduce to $E'$ (385.3 cm$^{-1}$) and $A'$$_1$ (404.5 cm$^{-1}$) modes from bulk to 1L, respectively.\cite{Anomalous_high_acsnano,Zhang_2015_csr,zhang-PhysRevB-2013} The ${E}_{2g}$-like mode shows an anomalous red-shift, instead of a blue-shift as for the  ${A}_{1g}$-like mode, from 1L to bulk, due to the dominance of long range Coulomb interaction.\cite{Anomalous_high_acsnano,Wirtz-phonons-prb} External perturbations, such as strain\cite{Strain-Raman-prb-mos2,Strain-phonon-prb-mos2etc}, temperature\cite{temperature-mos2-acsnano} and electric field\cite{Mos2-Raman-ele-prb}, can be used to tune the properties of 1L-MoS$_2$, and are reflected in $E'$ and $A'$$_1$ modes. Therefore, Raman spectroscopy is a useful tool to study optical properties of MoS$_2$ with external perturbations.\cite{Zhang_2015_csr}


\noindent {\bf \large Results}

Figure 1(a) shows the optical image of the MNRs with varied widths, from 5nm to 15 nm. The MNR arrays are patterned by direct helium ion beam milling (see Method section for details). These five nanoribbons arrays are patterned along the same direction with a single-crystalline region of a 1L-MoS$_2$ flake. Moreover, the atomic force microscope (AFM) is used to identify the width of the nanoribbons, as shown in Figure 1(b). The AFM image shows that the nanoribbon arrays are uniform with the equal spacing and width, as shown in Figure 1(c). The second harmonic generation (SHG) spectra (Figure 1(d)), which can be observed in the 1L-MoS$_2$ as an effective tool to identify the lattice orientation\cite{MoS2-shg-nl,shg-mos2-prb}, is employed to determine the ribbon crystal orientation. The intensity of SHG is proportional to sin$^2$$\phi$, where $\phi$ is the angle with respect to the zigzag orientation.\cite{MoS2-shg-nl} Thus, the SHG intensity would reach to the maximum when the laser beam is polarized along the armchair direction, which is $30^{\circ}$ with respect to the zigzag direction. As shown in Figure 1(d), the nanoribbon direction has an angle, about $7^{\circ}$, with respected to the zigzag orientation. The schematic diagram of the MNRs is shown in Figure 1(e) where the direction of the ribbon with respect to the MoS$_2$ crystal orientation is clearly indicated.


In the pristine 1L-MoS$_2$, the most prominent Raman peaks are $E'$ and $A'$$_1$ modes, which are at $\sim$384 and 405 cm$^{-1}$, respectively. Unlike black phosphorus\cite{bp-nn-xiaomu} or ReSe$_2$\cite{zhao-ReSe2-2015}, the in-plane cystal structure of 1L-MoS$_2$ is isotropic, therefore the Raman intensity of 1L-MoS$_2$ is independent of the polarization of the incident laser beam. However, it becomes strongly anisotropic when the 1L-MoS$_2$ is patterned into nanoribbons. Figure 2(a) plots the Raman spectra of the MoS$_2$ nanoribbons with 15nm width depending on linear polarization direction of the laser beam, the Raman spectra of MNRs with another widths and the pristine monolayer MoS$_2$ is shown in Fig. S2. As shown in Figure 2(b), the polarization angle of both the incident laser beam and scattered Raman signal are tuned by the same $\lambda$/2 plate, and the scattered light don't pass any polarizers after the $\lambda$/2 plate. We denote the angle between the polarization of the incident laser beam and the direction parallel to the nanoribbon as $\theta$. The polar plot of $E'$ and $A'$$_1$ peak intensity, I$_{rb}$($E'$) and I$_{rb}$($A'$$_1$), with respect to the $\theta$ is shown in Figure 2(c). Both I$_{rb}$($E'$) and I$_{rb}$($A'$$_1$) reach the maximum value when the incident laser beam is perpendicular to the ribbon direction, and when the incident laser beam is parallel to the ribbon direction the intensity become the minimum. It's different with the I$_{mono}$($E'$) and I$_{mono}$($A'$$_1$) ($E'$ and $A'$$_1$ intensity of monolayer MoS$_2$), which are the same for both direction. The Raman intensity in Figure 2(c) can be described as $a$${sin}^{2}$($\theta$)+$b$ ($a$ and $b$ are constant), which is similar to the behavior of the G mode in the graphene nanoribbons\cite{yang_2011_observation_NL}. Moreover,  ${I_{rb}}^{max}$($A'$$_1$)/${I_{rb}}^{min}$($A'$$_1$)$\sim$4, which is larger than that of the $E'$ modes ($\sim$2). The polarized Raman spectra are the direct evidence of the structural anisotropy of the MNRs.


Figure 3(a) shows the Raman spectra of the MNRs with different widths in the range of 120 - 480 cm$^{-1}$. In this measurement, the incident laser beam is perpendicularly polarized with respect to the nanoribbon direction. The Raman spectrum of the pristine 1L-MoS$_2$ is also plotted for comparison. In the pristine crystal, the phonon vibration can be considered as plane waves with finite wave vector $q\sim$0, due to the infinite spatial correlation. Therefore, the phonons at the center of Brillouin zone (BZ) can be detected by Raman scattering and the corresponding Raman mode shows a Lorentzian line shape. For a nanoribbon, there exists two edges, which will break the translational symmetry of the crystal. Because of the Heisenberg uncertainty principle, the fundamental $q \sim 0$ Raman selection rule is relaxed in MNRs, which make phonon uncertainty of $\Delta q\sim1/W$ be involved in the Raman scattering, where $W$ is the width of the nanoribbon. When $W$ become smaller, more phonons with different $q$ can be involved into the Raman scattering process. It will result in additional edge-activated Raman modes in the MNRs via the Heisenberg uncertainty principle. The contribution from each $q$ can be taken into account by a weighting function. A Gaussian function is found to agree with the experimental data.\cite{pfm-1986-Campbell} Therefore, for MNRs, there exist two categories of Raman modes: the intrinsic Raman modes near the $\Gamma$ point broadened by phonon confinement effect\cite{pfm-1981-Richter,pfm-1986-Campbell} and the edge-activated modes which result from the singularities in the vibrational density of states (VDOS) away from zone-center. The edge-activated modes always shows a Voigt (Gaussian-Lorentzian) profile. The line shape of intrinsic Raman modes in the MNRs can be described as\cite{pfm-1986-Campbell}:
\begin{equation}
I_{rb}(\omega )\propto \int \frac{exp(-{q}^{2}{W}^{2}/2\alpha) }{{\left[\omega -\omega (q) \right]}^{2}+{{\Gamma }_{0}}^{2}}dq
\label{eq:displ}
\end{equation}
\noindent where $\omega(q)$ describes the phonon dispersion, $\Gamma_0$ is the line width of Raman mode in bulk crystal, $\alpha$ is 8${\pi}^{2}$ and the integral is extended over the Brillouin zone. For the pristine 1L-MoS$_2$, $W$ tends to be infinity and eq(1) turns to a Lorentzian lineshape, whose center is at $\omega(0)$. This model has successfully explained the Raman scattering in the nanocrystals, such as ion-implanted graphene\cite{PhysRevB.82.125429}, graphite\cite{PhysRevB.49.1011} and 1L-MoS$_2$\cite{Mignuzzi_MoS2-defect}, microcrystalline silicon\cite{pfm-1981-Richter} and the silicon nanowires\cite{pfm-1986-Campbell}.

In Figure 3(a), the $E'$ and $A'$$_1$ modes are plotted using the blue solid lines, and the defect-activated modes are plotted using the green solid lines. At the low frequency range of 120-280 cm$^{-1}$, five Voigt peaks are used to fit the experimental data. The intensity of these low-frequency modes in the MNRs is much stronger than that of pristine 1L-MoS$_2$. These five sub-peaks are located at $\sim$155, $\sim$190, $\sim$210, $\sim$227 and $\sim$250 cm$^{-1}$, respectively. The peaks at $\sim$155 and $\sim$227 cm$^{-1}$ correspond to the two singularities in the VDOS, as shown in Figure 2(b), which is adapted from ref.\cite{Wirtz-phonons-prb}. The peak at 190 cm$^{-1}$ is linked to the shoulder on the right of the first singularity. As shown in the Figure s1, the two peaks at $\sim$155 and $\sim$190 cm$^{-1}$ can be assigned to ZA(M) (ZA phonon at M point) and TA(K), respectively. And the peak at $\sim$227 cm$^{-1}$ is assigned to the saddle point(SP) of the LA phonon branch (LA(SP)). The spectra in the range of 320-480 cm$^{-1}$ result from two categories of Raman modes as discussed above, the edge-activated modes and the intrinsic broadened modes. As shown in Figure 3(b), three prominent peaks are identified as the modes activated by defects, which correspond to TO(M) (357 cm$^{-1}$), LO(SP) (377 cm$^{-1}$) and ZO(M) (412 cm$^{-1}$), respectively. By adopting the phonon dispersion curve from Ref.\cite{Wirtz-phonons-prb}, the profile of $E'$ and $A'$$_1$ can be calculated using eq.(1). Combined with the three edge-activated modes, the experimental curves can be well fitted by the calculated results. Since the $E'$ mode is degenerated from TO and LO phonons, the contribution of these two phonon branches is also considered as a fitting factor.

As shown in Figure 3(a), I$_{rb}$($E'$) and I$_{rb}$($A'$$_1$) decrease with the reduction in nanoribbon width. Figure 3(c) shows I$_{rb}$($A'$$_1$) of the MNRs normalized with respect to ($I_p$($A'_1$)) of pristine 1L-MoS$_2$. The intensity ratio shows a linear trend with the ribbon width. Using a linear fitting, we can obtain a relationship between the intensity ratio and the ribbon width $W$: I$_{rb}$($A'$$_1$)/$I_p$($A'_1$)=0.0213$W$-0.0638. According to this relationship, the nanoribbon width can be determined by the measured I$_{rb}$($A'$$_1$)/$I_p$($A'_1$). It is extrapolated that the ratio will be close to unity when the width is $\sim$50 nm, indicating that 50nm width is the upper limit of this phonon confinement effect. Similarly, the fitting also gives the lower limit of the phonon-confinement effect is 3 nm under which I$_{rb}$($A'$$_1$) becomes zero. The line widths of $E'$ and $A'$$_1$ broaden with the decrease of the nanoribbon width, as shown in Figure 3(c). With decreasing the nanoribbon width, phonons with large $q$ would be involved, leading to a broader line width.


Figure 4(a) shows the Raman spectra of 1L-MNR with 15 nm width under different polarization configurations, the Raman spectra of MNRs with another widths and the pristine monolayer MoS$_2$ is shown in Fig. S3. For comparison, the Raman spectra of ion-implanted 1L-MoS$_2$ under the same polarization configurations are also plotted in Figure 4(b). As shown by the schematic diagram in Figure 4(a), the x-direction is perpendicular to the nanoribbon direction. Hence, the XX configuration means that the polarization of the incident laser beam is perpendicular to the ribbon direction, and the scattered Raman signal is collected along the same polarization direction. Similarly, the YX configuration means that incident laser beam is polarized parallel to the ribbon direction, and the scattered Raman signal is collected in the cross polarization direction. As shown in Figure 4(a), in the nanoribbon, the intensity of Raman peaks (both low-frequency and high-frequency modes) under the XX configuration is much stronger than that under the YY configuration. Moreover, the intensities under the cross configurations (XY and YX) is very similar to each other, which are weaker than that under XX configuration. Ion-implantation technique has been widely used to introduce defects in the bulk materials and obtain different domain sizes of nanocrystallines for graphite and multilayer graphenes \cite{tan-1999-APL-intrinsic,Ferreira-PRB-2010,QQli-Carbon-2015} and 1L--MoS$_2$.\cite{Mignuzzi_MoS2-defect} The defect-activated Raman modes will appear in ion-implanted materials because the small size of nanocrystallines make the fundamental $q \sim 0$ Raman selection rule relaxed. Indeed, when 1L-MoS$_2$ was implanted by Ar$^+$ with the dosage of 5$\times$10$^{12}$, besides the broadened $E'$ and $A'$$_1$ modes, several defect-activated Raman modes also appear in the range of 120-280 cm$^{-1}$, as reported in the previous studies.\cite{Mignuzzi_MoS2-defect} The ion-implanted 2D materials can be modeled as quantum dots with nano-sized diameter,\cite{Ferreira-PRB-2010} whose structures do not show any anisotropy, unlike the 1L-MoS$_2$ nanoribbons. Thus, the polarization-resolved Raman measurement on ion-implanted 1L-MoS$_2$ is expected to be very different from that of 1L-MoS$_2$ nanoribbons. Indeed, as shown in Figure 4(b), the intensities of the defect-activated modes and the $E'$ modes under different polarization configurations are almost identical to each other in ion-implanted 1L-MoS$_2$. This provides a direct evidence that the peculiar polarization behaviors of Raman modes in 1L-MoS$_2$ nanoribbons result from its structural anisotropy.

Optical contrast has been used to identify layer number of 2D materials on dielectric substrate,\cite{ni_2007_graphene_contrast} which can be measured by the reflection spectrum ($I_{2dm}(\lambda)$) from the flake region of 2D materials on the substrate and that ($I_{sub}(\lambda)$) from bare substrate not covered by 2D material flakes. The resulting optical contrast $\delta(\lambda)$ is then defined as $\delta(\lambda)$=($I_{sub}(\lambda)$-$I_{2dm}(\lambda)$)/$I_{sub}(\lambda)$. The optical contrast is so sensitive that the change of electronic band structures can be revealed from its optical contrast spectrum, as demonstrated in twisted multilayer graphenes\cite{wu-nc-2014-resonant}, WS$_2$ and WSe$_2$ flakes\cite{WJZhao-ACSnano-2013}. The optical contrast of MNRs with different widths and that of pristine 1L-MoS$_2$ under XX polarization are shown in Figure 5(a). The optical contrast of MNRs is quite different from that of pristine 1L-MoS$_2$, indicating a significant change of band structures after formation of MNRs. The optical contrast of MNRs is also very sensitive to its width. With decreasing ribbon width, the optical contrast is decreased. It suggests that narrower nanoribbons give the weaker absorption, which results from the more disordered overall crystal structure due to the defects created at the edges.

It is expected that the optical contrast of the pristine 1L-MoS$_2$ flake does not exhibit any anisotropy under XX and YY polarization configurations because of its isotropic structure. Indeed, in the inset of Figure 5(b), the optical contrast of the pristine 1L-MoS$_2$ is almost identical under XX and YY configurations. However, for the 15-nm MNR, the contrast under the YY configuration is smaller than that under the XX configuration within the measured spectrum range. The larger optical contrast of a 2D material flake means that the flake exhibits stronger absorption in the corresponding wavelength. It has been evident in the $\delta(\lambda)$ difference between AB-stacked and twisted bilayer (or four-layer) graphenes.\cite{wu-nc-2014-resonant} Therefore, the different $\delta(\lambda)$ between the two configurations indicates that 15-nm MNR exhibits bigger absorption for the incident light under the XX configuration than that under the YY configuration. The 5-nm MNR gives similar results, as demonstrated in Figure 5(b).

The optical contrast data discussed above is consistent with the result based on Raman spectroscopy. The Raman scattering is a third order process which involves the absorption of an incident photon, the creation (or annihilation) of a phonon, and the emission of a scattered photon. Thus, the intensity of Raman mode is proportional to $|$${M}_{eR}^{i}$$\cdot$${M}_{ph}$$\cdot$${M}_{eR}^{s}$$|$$^2$, where ${M}_{eR}^{i}$ and ${M}_{eR}^{s}$ is the optical transition matrix element (related to the adsorption) of the incident and scattered light, and ${M}_{ph}$ is the matrix element of the electron-phonon interaction. In 2D materials with reduced crystal symmetry, such as ReSe$_2$\cite{zhao-ReSe2-2015}, all the three terms contribute to the polarized Raman modes, and ${M}_{ph}$ plays a more dominant role. Here, we consider the intensity of $E'$ modes in 1L-MoS$_2$ nanoribbons, which is not sensitive to the polarization of the incident laser beam in pristine 1L-MoS$_2$, ${I}_{XX}(E')\propto\alpha|{M}_{eR}^{i}(X)\cdot{M}_{ph}(XX)\cdot{M}_{eR}^{s}(X)|^2$ and ${I}_{YX}(E')\propto\alpha|{M}_{eR}^{i}(Y)\cdot{M}_{ph}(YX)\cdot{M}_{eR}^{s}(X)|^2$. Larger absorption under the XX configuration leads to stronger Raman intensity. We find that ${I}_{XX}(E')$$\cdot$${I}_{YY}(E')$$\approx$${I}_{XY}(E')$$\cdot$${I}_{YX}(E')$. This indicates that ${M}_{ph}$ has no contribution to the polarization of the $E'$ mode, which is different from the anisotropic 2D materials with reduced symmetry, such as black phosphorus and ReSe$_2$.\cite{bp-nn-xiaomu,zhao-ReSe2-2015} Because the phonon vibration in MNRs is very similar to that in pristine 1L-MoS$_2$, it is reasonable that we observe the similar electron-phonon interactions in both of them.


\vspace*{5mm}
\noindent {\bf \large Conclusion}

In conclusion, All the Raman modes in MNRs show polarization-dependent intensity. The $E'$ and $A'$$_1$ peaks are broadened resulting from the phonon confinement effect with the reduction of nanoribbon widths. The relative intensity of $A'$$_1$ modes in MNRs to pristine 1L-MoS$_2$ can be used to probe the width of each nanoribbon. The edge of MNRs can activate additional new Raman modes, which correspond to the singularities in the VDOS. The polarization behavior of Raman modes is found to result from the anisotropy of the light absorption in the nanoribbons, which is supported by the optical contrast measurement. By patterning isotropic 2D materials, such as MoS$_2$, into nanoribbons, one can obtain high anisotropic optical properties.

\vspace*{5mm}
\noindent {\bf \large Methods}

\noindent {\bf Sample preparation.} The MNRs arrays were patterned by direct Helium ion beam milling using a Helium ion microscope (Orion plus, Carl Zeiss SMT GmbH). The energy of the Helium ion beam was about 30 KeV. The patterning was done using 10 m aperture size and beam spot 4, which gave 1.7 pA beam current. The nanoribbon patterns were formed by cutting arrays of single pixel lines with step size of 1 nm and dose of 5 nC/cm into the MoS$_2$ layer. The widths of the ribbons were controlled by the pitch of the cutting patterns.

\noindent {\bf Raman measurement.} Raman spectra are measured in back-scattering at room temperature with a Jobin-Yvon HR800 Raman system, equipped with a liquid-nitrogen-cooled charge-coupled device (CCD), a 100x objective lens (NA=0.90) and several gratings. The excitation wavelength is 488 nm from an Ar$^+$ laser. A 1800 lines/mm grating enables us to have the spectral coverage of 0.54cm$^{-1}$ at 488nm for each CCD pixel. The typical laser power is $\sim$0.5mW to avoid sample heating. The accumulation time for each spectrum is $\sim$600s.







\vspace*{5mm}
\bibliographystyle{naturemag}
\bibliography{MoS2_ribbon}

\newpage
\begin{figure*}[tb]
\centerline{\includegraphics[width=150mm,clip]{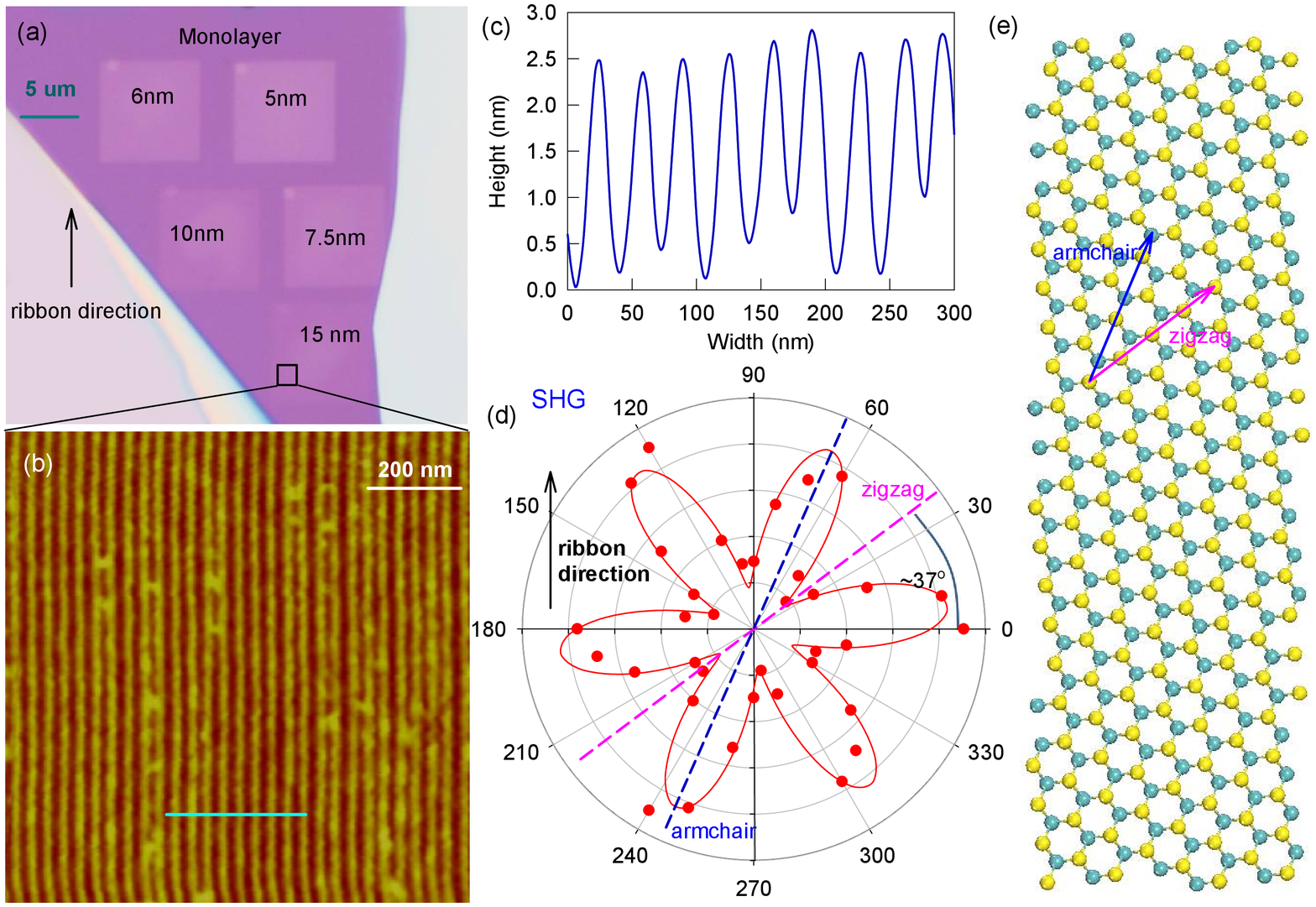}}
\end{figure*}
\textbf{Figure 1:}Characterization of the MoS$_2$ nanoribbons. (a) Optical image of 1L-MoS$_2$ nanoribbons. The nanoribbons with different width are milled in each 10$\times$10 um square. The ribbon direction is marked by the black arrow. (b) Atomic force microscopy image of MoS$_2$ nanoribbons with 15 nm width. (c) Height data with respect to the distance along cyan line marked in (b). (d) Second harmonic generation. The signal intensity depends on the incident laser polarization angle. The armchair and zigzag directions are marked by green and blue dash lines, respectively. (e) Schematic diagram of the MoS$_2$ nanoribbon. The armchar (A) and zigzag (Z) directions are marked by the blue and pink arrows, respectively.

\newpage
\begin{figure*}[tb]
\centerline{\includegraphics[width=120mm,clip]{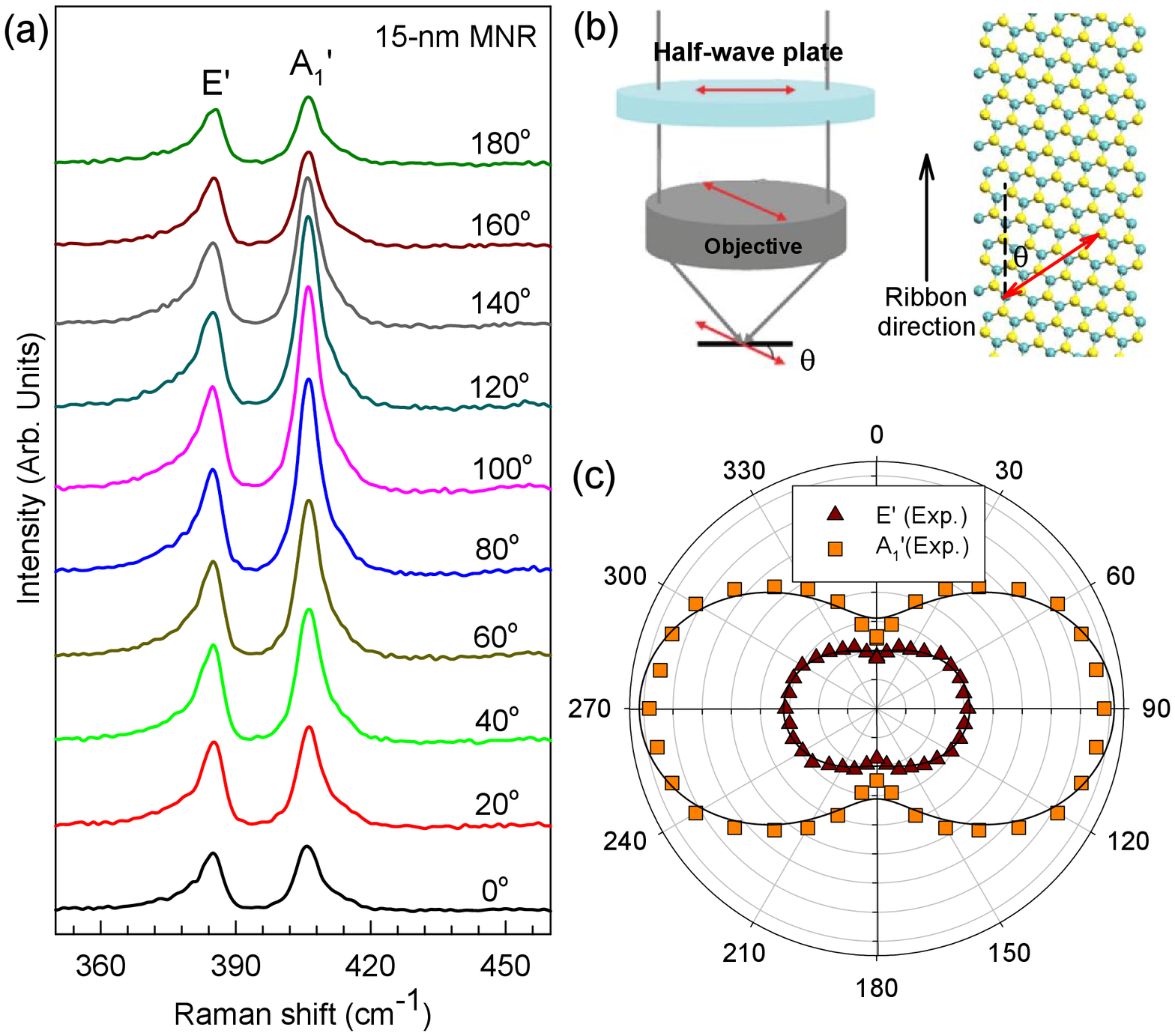}}
\end{figure*}
\textbf{Figure 2:} Polarized Raman spectra of the MoS$_2$ nanoribbon. (a) Raman spectra in the $E'$ and $A'$$_1$ peak region of the MoS$_2$ 15-nm nanoribbons with respect to the polarization angle of the incident laser beam. (b) Schematic diagram for the polarized Raman measurement, the half-wave plate is used for both incident and scattered light. The polarization angle $\theta$ is with respect to the direction of the nanoribbons. (c) Polar plot of the experimental $\theta$-dependent intensity of $E'$ and $A'$$_1$ modes.

\newpage
\begin{figure*}[tb]
\centerline{\includegraphics[width=150mm,clip]{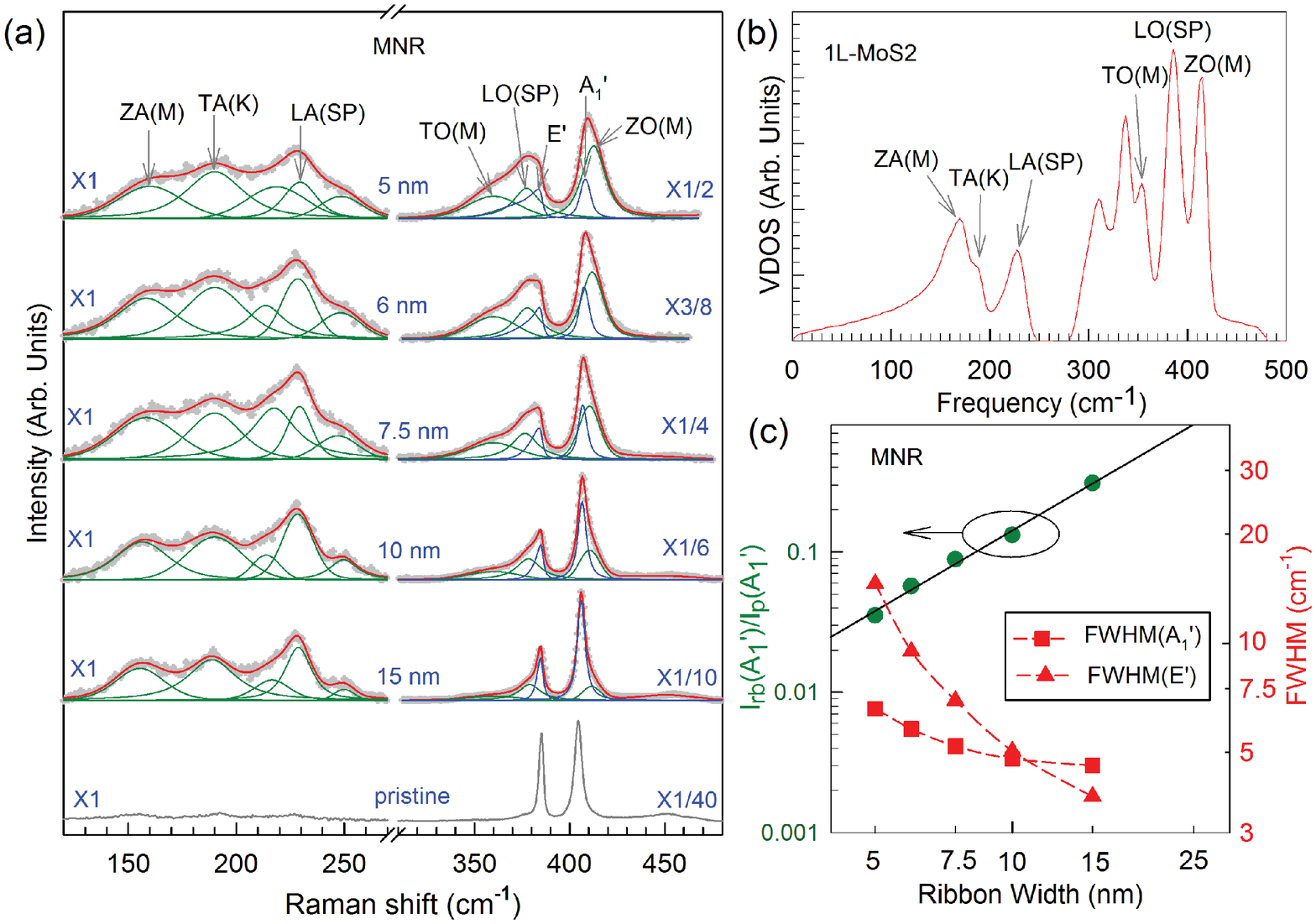}}
\end{figure*}
\textbf{Figure 3:} experimental data are shown by gray crosses, and the fitting data are plotted by red solid lines. The edge-activated modes and the broadened modes ($E'$ and $A'$$_1$) by quantum confinement effect are shown by green and blue solid lines, respectively. The Raman spectrum of pristine 1L-MoS$_2$ is plotted by gray solid line. The scaling factor of each spectrum is provided. (b) Vibrational density of states (VDOS) of 1L-MoS$_2$ from ref.\cite{Wirtz-phonons-prb}. The VDOS singularities of phonon branches at the high symmetry points of Brillouin zone are marked. (c) I$_{rb}$($A'$$_1$) of 1L-MoS$_2$ nanoribbons normalized by that (I$_{p}$($A'$$_1$)) of pristine 1L-MoS$_2$ with respect to the nanoribbon width, and full width at half maximum (FWHM) of the $E'$ and $A'$$_1$ modes with respect to the nanoribbon width.

\newpage
\begin{figure*}[htb]
\centerline{\includegraphics[width=150mm,clip]{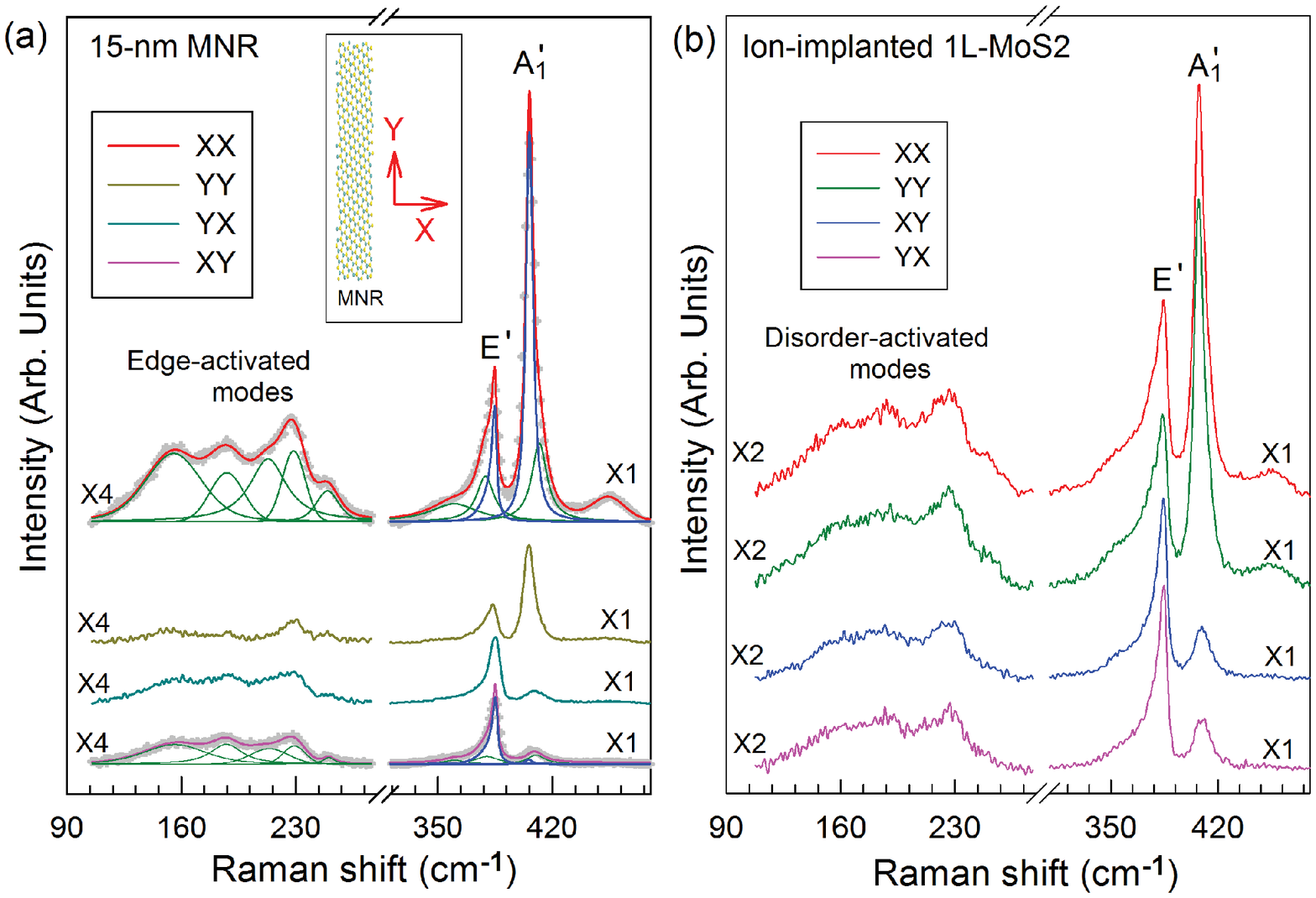}}
\end{figure*}
\textbf{Figure 4:} Polarized Raman spectra of MoS$_2$ nanoribbon and ion-implanted 1L-MoS$_2$. (a) Raman spectra of the 15 nm MoS$_2$ nanoribbons under different polarization configurations. For the XX and XY configurations, the defect-activated modes and the intrinsic $E'$ and $A'$$_1$ modes are shown by the green and blue solid lines, respectively. (b) Raman spectra of the ion-implanted 1L-MoS$_2$ under different polarization configurations.

\newpage
\begin{figure*}[tb]
\centerline{\includegraphics[width=150mm,clip]{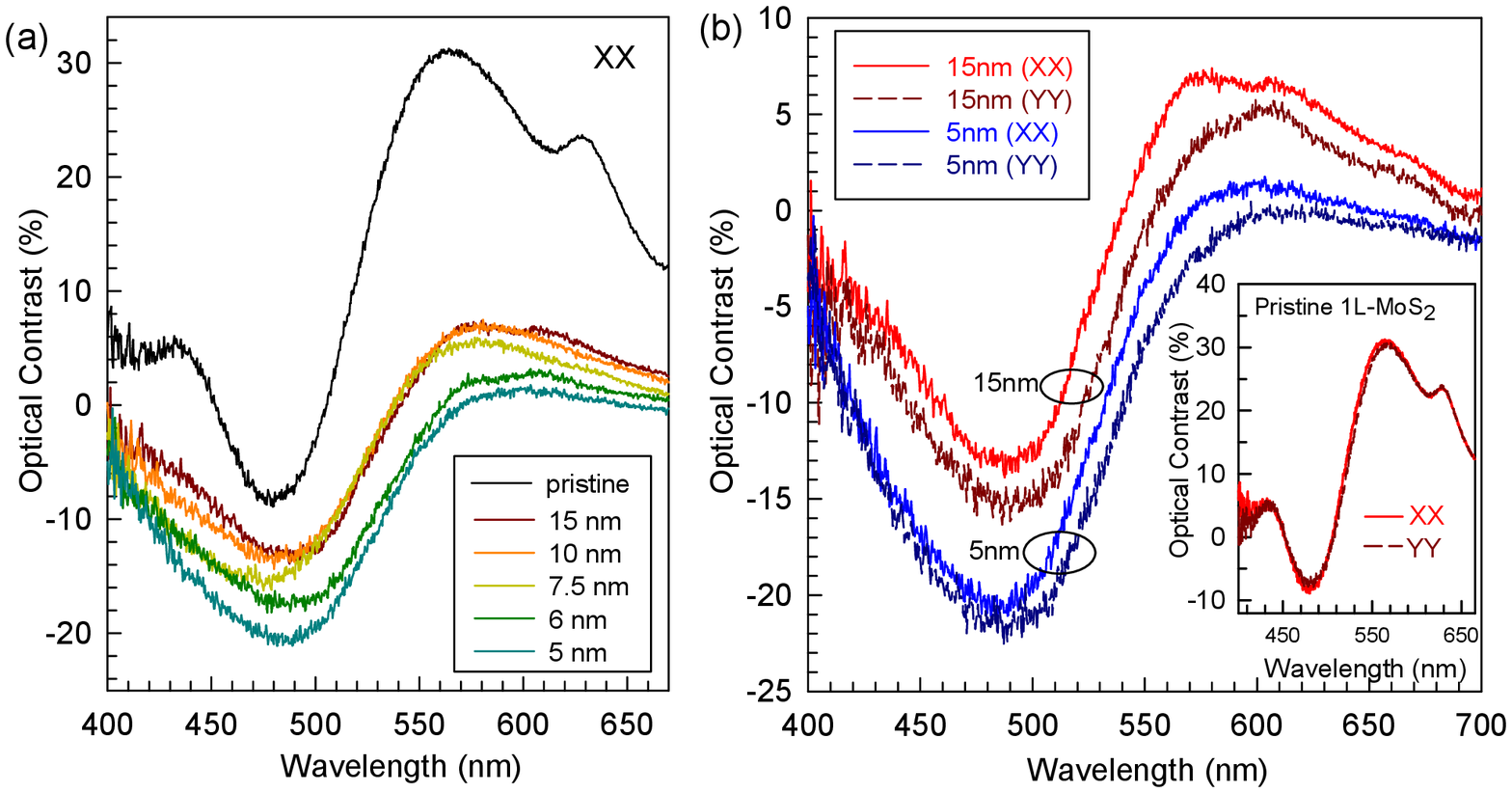}}
\end{figure*}
\textbf{Figure 5:} Optical contrast of MoS$_2$ nanoribbons with different width and polarization. (a) Optical contrast of MoS$_2$ nanoribbons with different width and that of pristine 1L-MoS$_2$ under XX polarizations. (b) Optical contrast of MoS$_2$ nanoribbons with 15 and 5 nm under different polarizations (XX and YY).  The contrast of monolayer MoS$_2$ with the same polarizations is shown in the inset.

\newpage
\textbf{The table of contents}\\
\textbf{The} monolayer MoS$_2$ is patterned into quasi-one-dimensional anisotropic MoS$_2$ nanoribbons (MNRs) with sub-20nm width. Systematic study of polarized optical contrast and Raman spectroscopy shows that the MNRs are with high optical anisotropy. The study paves the way to explore quasi-one-dimensional materials with high optical anisotropy from isotropic 2D family of transition metal dichalcogenides.\\

{\bf Keywords}: Molybdenum Disulfide, Nanoribbon, Raman spectroscopy, Anisotropy.\\

Jiang-Bin Wu$^1$, Huan Zhao$^1$, Yuanrui Li$^1$, Douglas Ohlberg, Wei Shi, Wei Wu*, Han Wang*, Ping-Heng Tan*\\

\textbf{Monolayer Molybdenum Disulfide Nanoribbons with High Optical Anisotropy}\\

\begin{figure*}[htb]
\centerline{\includegraphics[width=110mm,clip]{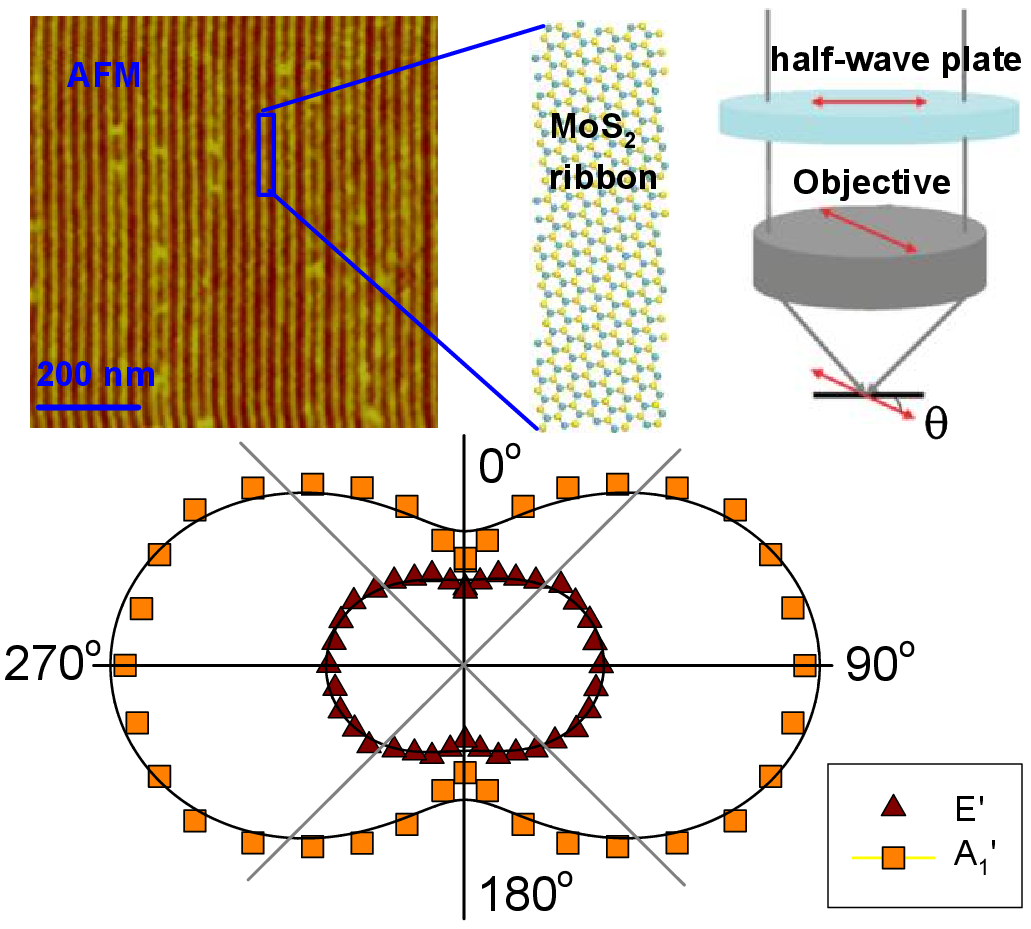}}
\end{figure*}

\newpage
\textbf{Supporting Information}\\
\textbf{Monolayer Molybdenum Disulfide Nanoribbons with High Optical Anisotropy}\\
Jiang-Bin Wu$^1$, Huan Zhao$^1$, Yuanrui Li$^1$, Douglas Ohlberg, Wei Shi, Wei Wu*, Han Wang*, Ping-Heng Tan*\\

\begin{figure*}[!]
\centerline{\includegraphics[width=120mm,clip]{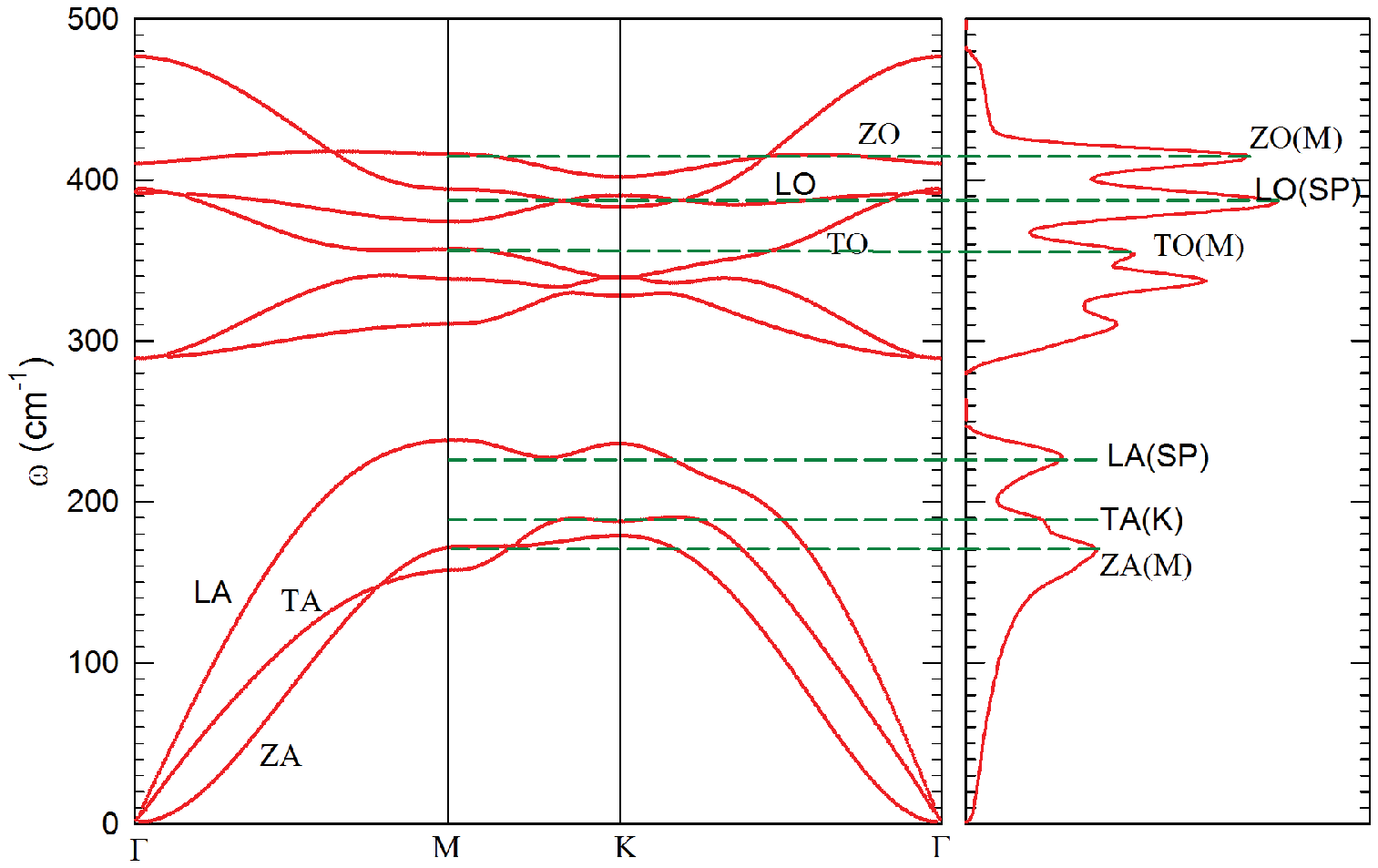}}
\end{figure*}
\textbf{Figure S1: Phonon dispersion and vibration density of states of monolayer MoS$_2$.} The singularities in the vibration density of states (VDOS)are assigned to the phonon branch at the Brillouin zone edge or the saddle points. Phonon dispersion and VDOS are from ref.25.

\newpage
\begin{figure*}[!]
\centerline{\includegraphics[width=120mm,clip]{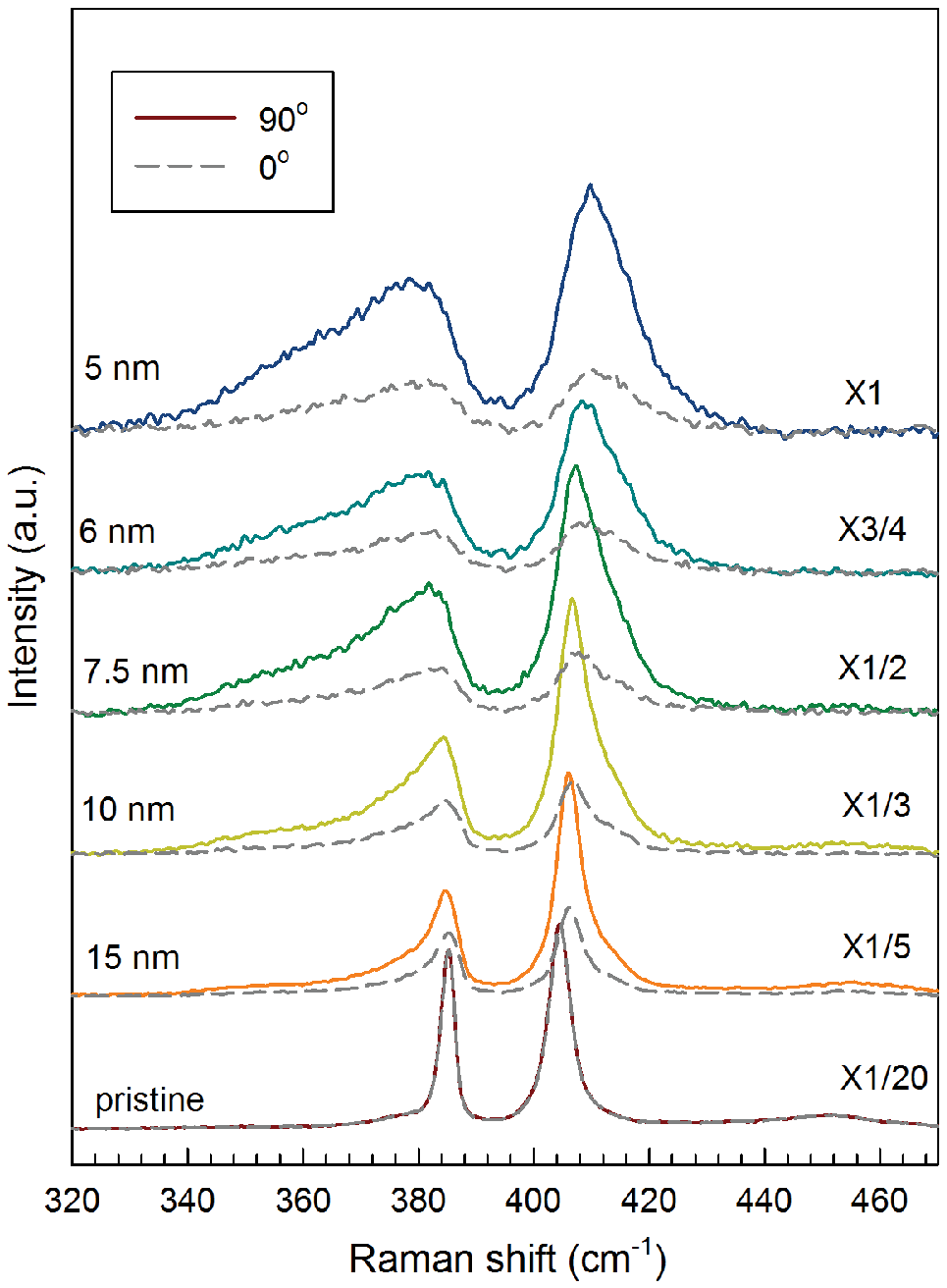}}
\end{figure*}
\textbf{Figure S2: Polarized Raman spectra of the MoS$_2$ nanoribbon with different widths.} The polarization configure is the same with that in Fig.2.
\newpage
\begin{figure*}[!]
\centerline{\includegraphics[width=120mm,clip]{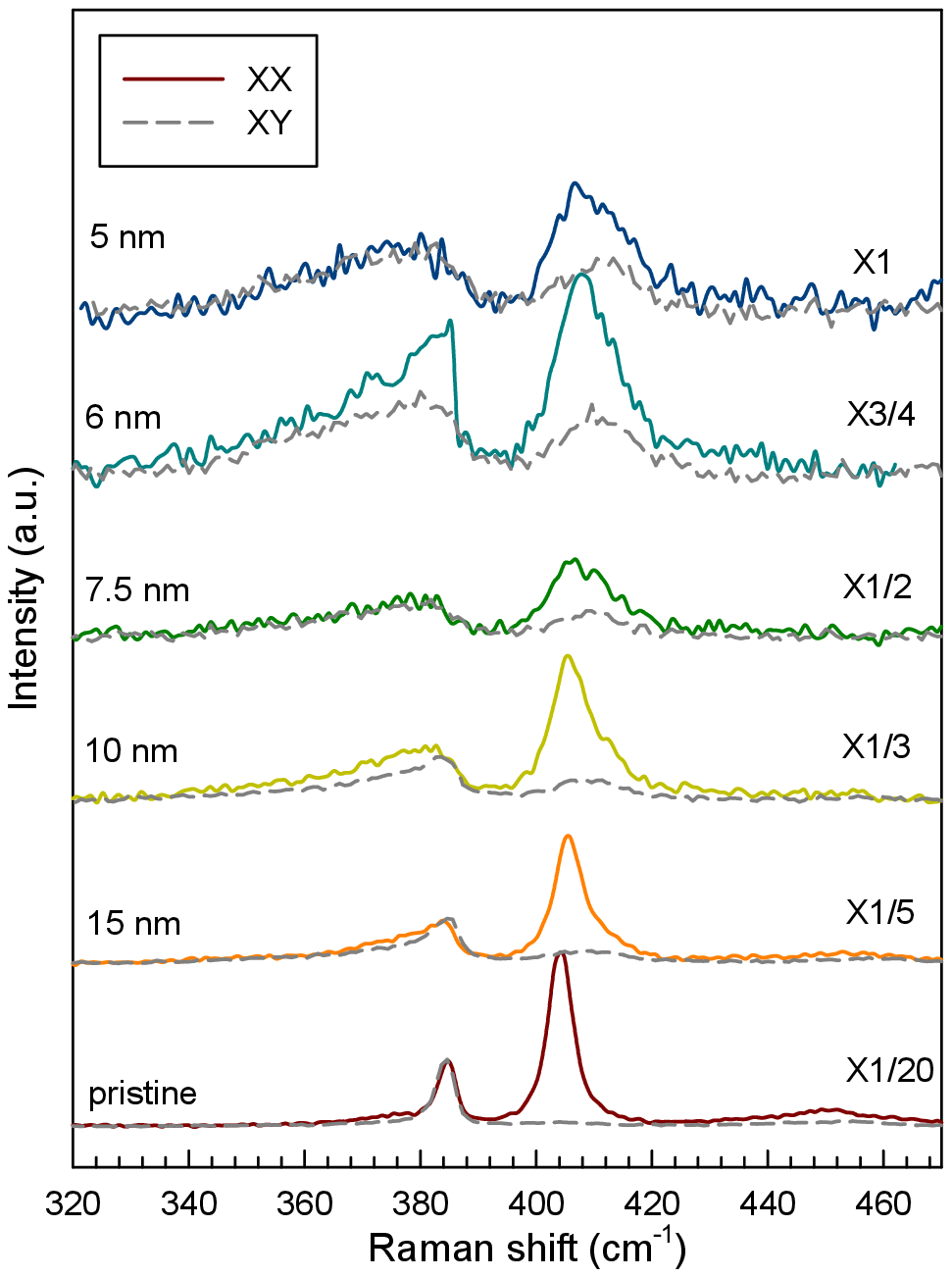}}
\end{figure*}
\textbf{Figure S3: Polarized Raman spectra of the MoS$_2$ nanoribbon with different widths.} The polarization configure is the same with that in Fig.4.

\end{document}